\documentstyle[12pt]{article}
\let\LARGE=\Large
\let\Large=\large
\let\large=\normalsize
\newcommand{\be}[3]{\begin{equation}  \label{#1#2#3}}     
\newcommand{\ee}{ \end{equation}}
\newcommand{\ba}{\begin{array}}
\newcommand{\ea}{\end{array}}

\newcommand{\PRL}[2]{{\em Phys. Rev. Lett.}{ \bf #1#2}}

\newcommand{\PL}[3]{{\em Phys. Lett.}{ \bf B#1#2#3}}

\begin{document}
\thispagestyle{empty}
\rightline{HUB-EP-97/30}
\rightline{hep-th/9705169}
\vspace{2truecm}
\centerline{\bf \LARGE
Stationary solutions of $N=2$ supergravity}
\vspace{1.2truecm}
\centerline{\bf 
Klaus Behrndt\footnote{behrndt@qft2.physik.hu-berlin.de}\ , \ 
Dieter L\"ust\footnote{luest@qft1.physik.hu-berlin.de}
 \  and  \  Wafic A. Sabra\footnote{sabra@qft2.physik.hu-berlin.de}}
\vspace{.5truecm}
{\em 
\centerline{Humboldt-Universit\"at, Institut f\"ur Physik}
\centerline{Invalidenstra\ss e 110, 10115 Berlin, Germany}
}

\vspace{1.2truecm}
\vspace{.5truecm}
\begin{abstract}
We discuss general bosonic stationary configurations
of $N=2$, $D=4$ supergravity coupled to vector multiplets. 
The requirement of unbroken
supersymmetries imposes constraints on the holomorphic symplectic section of 
the underlying 
special K\"ahler manifold.
The corresponding solutions of the field equations are completely
determined by a set of harmonic functions. As examples we
discuss rotating black holes, Taub-NUT and Eguchi-Hanson like instantons
for the $STU$ model. In addition, we discuss, in the static limit, 
worldsheet instanton corrections to the $STU$ black hole solution, 
in the neighbourhood of a vanishing 4-cycle of the Calabi-Yau manifold. 
Our procedure is quite
general and includes all known black hole solutions that can
be embedded into $N=2$ supergravity.
\end{abstract}
\bigskip \bigskip
\newpage

\section{Introduction} 

During the recent years there has been an enormous progress in the
understanding of non-perturbative phenomena in $N=2$ supersymmetric
field theories \cite{SW} as well as in $N=2$ string vacua in
four-dimensions. In particular, strong-weak coupling and string/string
duality symmetries \cite{KV} among type II compactifications on
Calabi-Yau threefolds and heterotic vacua on $K3\times T^2$ with $N=2$
supersymmetry in four dimensions were successfully tested for a class 
of models \cite{test}.  Moreover, various types of
(non)-perturbative transitions among different $N=2$ string vacua were
explored.  Because of all this progress, the unification of possibly
all superstring theories can very likely be realized by common
underlying framework, often called $M$-theory.

Transitions among $N=2$ string vacua generically can take place at
those points in the moduli spaces where non-perturbative BPS states
become massless.  For example, considering $N=2$ type II string vacua
on Calabi-Yau three-folds, electrically or magnetically charged black
holes become massless at the conifold points in the Calabi-Yau moduli
spaces, where certain homology cycles of the Calabi-Yau spaces shrink
to zero size.  Therefore it is very important to find the most general
solutions of the effective $N=2$ supergravity action coupled to $N=2$
matter multiplets.  By constructing non-trivial space-time dependent
solutions of $N=2$ supergravity one can hope to obtain more
information about the dynamical nature of the non-perturbative
transitions. For instance, for generic non-trivial black hole
solutions, not only the four-dimensional metric but also the
four-dimensional gauge fields as well as the moduli fields are
expected to be space-time dependent functions. Then one deals with
internal Calabi-Yau spaces whose shapes vary over every point in the
four-dimensional space-time; in this way one gets a very interesting
link between the structures of the internal space and of
four-dimensional space-time. In particular, one can try to determine
those points in space-time where the internal Calabi-Yau periods
shrink to zero size and where the transitions due to massless black
holes take place.  Therefore, the black hole solutions may provide an
interesting link between external and internal singularities.

In this paper we will study stationary (i.e.  time indepedent)
solutions of $N=2$ supergravity coupled to $N=2$ vector multiplets.
In recent times there has been a considerable progress in the
understanding of extreme solutions of $N=2$ supergravity in four
dimensions.\footnote{A recent discussion of non-extreme Calabi-Yau
black holes appeared in \cite{KaWin}.}  In contrast to $N=4,8$
theories, where the dynamics is highly restricted by supersymmetry,
for $N=2$ solutions one may expect a much richer structure. E.g.\
black holes of $N=4,8$ supersymmetric theories are expected not to
receive quantum corrections whereas solutions of $N=2$ supergravity
will alter significantly when quantum corrections are taken into
account. This opens the possibility to address, besides the already
mentioned problems, many interesting questions: Are black holes
stable? What happens with the singularity? Are quantum correction
encoded in the Hawking temperature?

To find general solutions that include quantum corrections is in
general a difficult problem - to solve the equations of motion of
quantum gravity seems to be hopeless. One 
interesting first starting point is to look at the
entropy of non-singular static black holes.  
Here, a major breakthrough
came with the observation, that the entropy can be obtained by
extremizing the central charge of the underlying $N=2$ super-algebra 
with respect to the moduli
\cite{fe/ka1}, \cite{fe/ka2}. It was also shown that the scalar fields
follow attractor equations with fixed points on the horizon. These
fixed points are independent of the moduli (scalar fields) at infinity
and are determined by the conserved charges and topological data
only. If the scalar fields take this value throughout
the entire space-time, i.e.\ they are
constant, one gets the so-called double extreme black holes \cite{ka/sh}.

All these results are covariantly formulated in terms of the underlying 
special geometry, i.e., all quantities can be expressed in terms of the 
symplectic sections. The central statement is that the symplectic section 
has to fulfil
constraints, the so-called stabilisation equations.  
Specifically, these equations are satisfied if the symplectic vector
$\Pi=(Y^I,M_I(Y))$, where $Y^I=\bar Z L^I,$ with $Z$ being the $N=2$ 
central
charge and $(L^I,M_I)$ being the symplectic period vector of $N=2$
supergravity (covariantly holomorphic section), fulfils the following set 
of equations \cite{be/ca}
\be009
i (Y^I - \bar{Y}^I ) = p^I \qquad , \qquad i (F_I(Y) -\bar{F}_I(\bar Y)) =
q_I \ ;
\ee
here the $p^I$ ($q_I$) are the integer valued magnetic (electric)
charge vectors.
Solutions of
these equations determine the double extreme black holes completely or
the entropy for any static black hole with a non-singular horizon.
For the double extreme black holes, the  equations (\ref{009}) are
also called ``stabilisation equations''.  The motivation was the
following. The scalar fields at infinity take constant values
(moduli), which are not protected by any kind of gauge
symmetry. Therefore, one can argue that dynamically a given
configuration will ``choose'' those values for which the mass
(=central charge) becomes minimal \cite{re}.  As long as the solution is
non-singular one can show that for these values the scalar fields
become constant everywhere and these constants ensure that the mass is
minimal. The solution (black hole) has been stabilised. 
By means of the equations (\ref{009}) it is straightforward to 
compute the classical entropy of static $N=2$ 
black holes \cite{be/ka1}, \cite{sa1} as well as to incorporate quantum
corrections \cite{be/ca}, \cite{be/ga}. 
So the full entropy depends in the heterotic case
on the space-time instanton numbers, whereas in the type II compactifications
the entropy depends on the topological data of the Calabi-Yau 
spaces, like the intersection numbers, the Euler number and the rational 
worldsheet instanton numbers.

In the meanwhile, these results have been extended in several
directions. 
For example, extreme solutions with non-constant scalar fields were 
constructed \cite{Klaus} for the axion-free $STU$ model with cubic
prepotential, and for supergravity models based on quadratic 
prepotentials \cite{sa1}. 
In these 
models it was found that the static solutions are expressed in terms of 
constrained harmonic functions. Later it was in fact shown \cite{sa2}
that for general static extreme
$N=2$ black holes, the solutions are completely specified
by the K\"ahler potential of the underlying special geometry where the 
imaginary part of the holomorphic sections are given in terms of a 
set of constrained harmonic functions.
In addition, for certain cases a microscopic understanding of the
$N=2$ entropy formulae was achieved by wrapping various branes around
the internal Calabi-Yau cycles \cite{km}.

The aim of the present paper is to discuss a generalisation 
of this approach, which can be used for any static or stationary $N=2$
solution. It will be shown, that a supersymmetric configuration is given,
if the symplectic holomorphic section $\Omega =(X^I , F_I)$ 
satisfies the equations
\be010
i (X^I - \bar{X}^I ) = \tilde{H}^I(x^{\mu}) \qquad , \qquad 
i (F_I -\bar{F}_I) = H_I(x^{\mu}) 
\ee
where the symplectic vector $\left(\tilde{H}^I(x^{\mu}) , H_I(x^{\mu})
\right)$
defines the gauge fields. In addition, imposing the equations of
motion and Bianchi identities these functions have to be harmonic.
These equations can be seen as a consistent bosonic truncation of
$N=2$ supergravity, off-shell if the $H's$ are arbitrary and on-shell
for harmonic $H's$.

The above equations (\ref{010}) are very interesting, since they
combine the structure of the internal space, e.g.\ a Calabi-Yau
threefold, on the left hand side with space-time properties on the
right hand side. Consequently, as discussed at the beginning,
singularities in space time are related to special points in the
internal space, like walls of the K\"ahler cone or conifold
singularities.

Moreover, the equations (\ref{010}) provide black hole solutions for
cases where the horizon is singular. But for these singular solutions,
the scalar fields cannot be stabilised. Again, one would expect that
the moduli will dynamically arrange each other in such a way that the
mass becomes minimal. But this minimum is {\em not} described by the
equations (\ref{010}). Instead, these equations describe the complete
supersymmetric configuration, in general not with an extremized
central charge. Therefore, they open especially the possibility to
investigate singularities, e.g.\ for topology change of the internal
space (all kinds of phase transitions). In addition one can discuss
solutions, that have not enough charges for a non-singular horizon, or
solutions which do not have any horizon like Taub-NUT spaces, or naked
singularities like for supersymmetric rotating black holes. Note, that
the equations (\ref{010}) allow to include all perturbative and
non-perturbative corrections - only higher derivative corrections are
not (yet) taken into account.

As already said, the $N=2$ black hole solutions constructed so far,
describe static configurations. In this case the metric depends on the
$N=2$ K\"ahler potential. We will show that stationary, but non-static
solutions are obtained if the metric also depends on the
$U(1)$ K\"ahler connection, which is also a symplectic invariant
quantity in the context of $N=2$ special geometry.
Depending on the choices for the harmonic functions and on the
considered prepotentials, one gets 
for example non-static
rotating $N=2$ black holes,  $N=2$ Taub-NUT spaces or $N=2$ 
Eguchi-Hanson
like instantons. Note that similar solutions were also discussed previously
in the context of $N=2$ supersymmetric non-linear $\sigma$-models
\cite{sigma}.

The paper is organised as follows.
In the next chapter we will collect some formulae and expressions
of $N=2$ supergravity which will be important for the following discussion.
In section three we will determine the stationary solutions by the
requirement that half of the $N=2$ supersymmetries are unbroken
in the considered background. In addition, the field equations constrain
the functions, which appear in the solutions of the gravitino and
gaugino variations, to be harmonic. Then, in the fourth chapter
we will apply our formalism to construct  examples of stationary
$N=2$ solutions. Specifically, after considering possible choices
for the  harmonic functions, we will 
explicitly discuss the case of pure supergravity, and 
rotating black holes, Taub-NUT spaces and
Eguchi-Hanson instantons in the so-called $STU$ model.
We also discuss worldsheet instanton corrections to the 
static $STU$ black hole solution, in the neighbourhood of a vanishing 4-cycle 
of the Calabi-Yau manifold. 
Some conclusions will close the paper. Our conventions are collected in the
appendix.

\section{Special geometry and $N=2$ supergravity}

The structure of $N=2$ supergravity theories coupled to vector and
hypermultiplets is governed by special geometry. In this section, we
briefly review some of the essential formalism of special geometry
which will be relevant for our discussion.  The complex scalars $z^A$
of the $N=2$ vector multiplets coupled to supergravity are coordinates
which parametrise a special K\"ahler manifold. This is a
K\"ahler-Hodge manifold, with an additional constraint on the
curvature \cite{sspecial}
\be012
R_{A \bar B C \bar D}=g_{A \bar B} \, g_{C \bar D}+g_{A \bar D} \,
g_{C \bar B} - C_{ACE}\, C_{\bar B \bar D \bar L} \, g^{E \bar L},
\ee
where $g_{A\bar B }=\partial_A\partial_{\bar B} K,$ is the K\"ahler
metric with $K$ the K\"ahler potential and $C_{ABC}$ is a completely
symmetric covariantly holomorphic tensor. K\"ahler-Hodge manifolds are
characterised by a $U(1)$ bundle whose first Chern class is equal to
the K\"ahler class.  This implies that, locally, the $U(1)$ connection
can be represented by
\be014
Q=-{i\over2}(\partial_A K dz^A-\partial_{\bar A}K d\bar z^{A}) \ .
\ee
An intrinsic definition of special K\"ahler manifold can be given
\cite{c}-\cite{cccc} in terms of a flat $2n+2$ dimensional symplectic
bundle over the K\"ahler-Hodge manifold, with the covariantly
holomorphic sections
\be016
\ba{l}
V=\pmatrix{L^I\cr M_I}, \qquad I=0,\cdots,n \\
D_{\bar A} V = (\partial_{\bar A}- {1\over2}\partial_{\bar A}K)V=0,
\ea
\ee
obeying the symplectic constraint
\be018
i\langle V\vert\bar V \rangle =i (\bar L^I M_I-L^I\bar M_I)=1
\ee
where the symplectic inner product is understood to be taken with
respect to the metric $\pmatrix{0&-1\cr 1&0}.$ One also defines
\be020
U_A=D_A V=(\partial_{A} + {1\over2}\partial_{A}K)V=
\pmatrix{f_A^I\cr h_{AI}}.
\ee
In general, $D_A$ is the covariant derivative with respect to the
Levi-Civita connection and the connection $\partial_A K$. Thus, for a
generic field $\phi^i$ which transforms under the K\"ahler
transformation, $K\rightarrow K+f+\bar f$, by the $U(1)$
transformation $\phi^A \rightarrow e^{-({p\over2}f+{{\bar
p}\over2}{\bar f})}\phi^A,$ we have
\be022
D_A\phi^B=\partial_A\phi^B+\Gamma^B_{{AC}}\phi^C+{p\over2}{\partial_A 
K}\phi^B.
\ee
One also defines the covariant derivative $D_{\bar A}$ in the same
way but with $p$ replaced with $\bar p.$ The sections $(L^I, M_I)$ has
the weights $p=-\bar p=1$ and $C_{ABC}$ has the weights $p=-\bar p=2$.

In general, one can write
\be024
\ba{l}
M_I={\cal N}_{IJ}L^J \ ,   \\
h_{AI}={\bar{\cal N}}_{IJ}f_{A}^J \ .
\ea
\ee
The complex symmetric $(n+1)\times (n+1)$ matrix $\cal N$ encodes the
couplings of the vector fields in the corresponding $N=2$ supergravity
theory.

It can be shown \cite{sspecial}-\cite{special} that the condition
(\ref{012}) can be obtained from the integrability conditions on
the following differential constraints
\be026
\ba{l}
D_AV = U_A \ , \\
D_AU_B =  iC_{ABC}g^{C \bar L }{\bar U}_{\bar L } \ , \\
D_A{\bar U}_{\bar B } =  g_{A \bar B}{\bar V} \ , \\
D_A{\bar V} = 0 \ ,
\ea
\ee
and \cite{v}
\be036
\langle V, U_A \rangle =0.
\ee
It is well known that the above constraints can in general be
solved in terms of a holomorphic function of degree two
\cite{sspecial}. However, there exists symplectic sections for which
such a holomorphic function does not exist. This, for example, appears
in the study of the effective theory of the $N=2$ heterotic strings
\cite{CDFP}. Thus it is more natural to use these differential
constraints as the fundamental equations of special
geometry.

The K\"ahler potential can be constructed in a symplectic invariant
manner as follows. Define the sections $\Omega$ by
\be028
V=\pmatrix{L^I\cr M_I}=e^{K\over2}\Omega=e^{K\over2}\pmatrix{X^I\cr 
F_I}\ .
\ee
It immediately follows from (\ref{016}) that $\Omega$ is
holomorphic;
\be030
\partial_{\bar A}X^I=\partial_{\bar A}F_I=0 \ .
\ee
Using ({\ref{018}), one obtains 
\be032
\ba{rcl}
K&=&-\log\Big(i\langle \Omega\vert\bar\Omega \rangle \Big) \\
 &=&-\log\Big[i(\bar X^IF_I-X^I\bar F_I)\Big].
\ea
\ee
Exploiting the relations (\ref{018}),(\ref{026}) and (\ref{024}),
the following symplectic expression can be obtained for the K\"ahler metric
\be034
g_{A\bar B}=-i \langle U_A \vert\bar U_{\bar B }\rangle =
-2f_A^I \mbox{Im}{\cal N}_{IJ}\bar f_{\bar B}^J.
\ee
For our purposes, it is also useful to display the following relations 
\be038 
g^{A \bar B} f_A^I{\bar f}_{\bar B}^J=- {1\over2}(\mbox{Im}
{\cal N})^{IJ} - {\bar L}^IL^J \ .
\ee
and 
\be570
F_I \partial_{\mu} X^I - X^I \partial_{\mu} F_I =0 \ .
\ee
which is a consequence of (\ref{036}).

It should be mentioned that the dependence of the gauge couplings on
the scalars characterising homogenous special K\"ahler manifolds of
$N=2$ supergravity theory can also be determined from the knowledge of
the corresponding embedding of the isometry group of the scalar
manifold into the symplectic group \`a la Gaillard and Zumino
\cite{gz, sym}.

The $N=2$ supergravity action includes one gravitational, $n$ vector
and hypermultiplets. However, for our purposes, the
hypermultiplets are assumed to be constants. 
In this case, the bosonic $N=2$ action is 
given by
\be040
S_{N=2}= \int \sqrt{-g} \, d^4x \Big(-{1\over2}R + g_{A \bar B}
\partial^{\mu} z^A \partial_{\mu} \bar z^{B} + i \left(\bar
{\cal N}_{I J} {F}^{- I}_{\mu \nu} {F}^{- J {\mu \nu}} \, - \, {\cal
N}_{IJ} {F}^{+ I}_{\mu \nu} { F}^{+ J {\mu \nu}}\right)
\ee
where 
\be042
{F}^{\pm I \, \mu\nu} = {1\over2}\Big({F}^{I \, \mu\nu} \pm 
{i\over2}\varepsilon^{\mu\nu\rho\sigma}{F}_{\rho\sigma}^I\Big).
\ee
The important field strength combinations which enter the chiral
gravitino and gauginos supersymmetry transformation rules are given by
\be044
\ba{rcl}
T^{-}_{\mu\nu}&=& M_I F^I_{\mu\nu}-L^I G_
{I \mu\nu}=2i(\mbox{Im} {\cal N}_{IJ})L^I F^{J-}_{\mu\nu}\\
G_{\mu\nu}^{-A}&=&-g^{A \bar B}{\bar f_{\bar B}^I} (\mbox{Im}{\cal 
N})_{IJ}F^{J-}_{\mu\nu}
\ea
\ee
with
\be130
G_{I\, \mu\nu} = \mbox{Re} {\cal N}_{IJ} F^{J}_{\mu\nu} - 
 \mbox{Im} {\cal N}_{IJ} {^{\star}}F^J_{\mu\nu} \ .
\ee

The supersymmetry transformation for the chiral gravitino
$\psi_{\alpha\mu}$ and gauginos $\lambda^{A \alpha}$ in a bosonic
background of $N=2$ supergravity are given by
\be046
\ba{rcl}
\delta\,\psi_{\alpha\mu} &=& \nabla_\mu \epsilon_\alpha -
\frac{1}{4} T^-_{\rho\sigma} \gamma^{\rho} \gamma^{\sigma}
\, \gamma _\mu \, \varepsilon_{\alpha\beta}\epsilon^\beta \, 
\label{grtrans},\\
\delta\lambda^{A\alpha} & = & i \, \gamma^\mu \partial_\mu
z^A \epsilon^\alpha + G^{-A}_{\rho\sigma}\gamma^{\rho}\gamma^{\sigma}
\varepsilon ^{\alpha\beta} \epsilon_\beta
\ea
\ee
where $\epsilon_\beta$ is the chiral supersymmetry parameter,
$\varepsilon^{\alpha\beta}$ is the $SO(2)$ Ricci tensor and the
space-time covariant derivative $\nabla_{\mu}$ also contains the
K\"ahler connection
\be048
Q_\mu= -{i\over2}\Big(\partial_A K\partial_\mu z^A-
\partial_{\bar A } K\partial_\mu {\bar z}^{A}\Big),
\ee
Therefore we have 
\be050
\nabla_\mu \epsilon_\alpha=(\partial_\mu-{1\over4}w^{ab}_\mu\gamma_a 
\gamma_b + {i\over 2}Q_\mu)\epsilon_\alpha 
\ee
where $w^{ab}_\mu$ is the spin connection. 
\section{Stationary solutions}
In this section we will first describe the stationary solution. In a
second and third subsection we show that this solution is
supersymmetric, i.e.\ the gravitino and gaugino variations vanish.
Since the hyperscalars are trivial in our model, the hyperino
variation is identically fulfilled.  We will see that supersymmetry
does not restrict the functions $(\tilde H^I, H_I)$ to be
harmonic. This property is a consequence of the equations of motions and the
Bianchi identities of the gauge fields.
\subsection{The solution}
First, we investigate the gauge field part, which can be written as
\be120
\int ( \mbox{Im} {\cal N}_{IJ} F^I \cdot F^J + \mbox{Re}
 {\cal N}_{IJ} F^{I} \cdot {^{\star}F^{J}})  = \int F^I {^{\star}}G_I
\ee
where we have ignored space-time indices, i.e.,\ $F^I \cdot F^J
\equiv F^I_{\mu\nu} F^{J \, \mu \nu}$ and $G_{I \, \mu\nu}$
is defined in (\ref{130}).
The equations of motion and the Bianchi identities for the gauge fields are 
given by
\be140
\partial_{\mu} (\epsilon^{\mu\nu\rho\lambda} G_{I\, \rho\lambda}) =0
\quad , \quad \partial_{\mu} (\epsilon^{\mu\nu\rho\lambda} 
F^I_{\rho\lambda}) = 0 \ .
\ee
Assuming that our solution is time independent, the spatial components are
given by
\be150
F^I_{mn} = {1 \over 2} \epsilon_{mnp} \partial_p \tilde{H}^I
\quad , \quad G_{I\, mn} = {1 \over 2} \epsilon_{mnp} \partial_p H_I
\ee
where $m,n,p= 1,2,3$ and $(\tilde{H}^I , H_I)$ are harmonic functions.
Note, that the timelike components $F^I_{0m}$ and $G_{I\, 0m}$ are 
fixed in terms of the spatial ones using 
(\ref{130}).
As next step we consider the metric. We are interested in time
independent supersymmetric solutions. It has been shown by Tod \cite{to}
that this requires an IWP metric \cite{is/wi} which can be expressed in the 
following form
\be160
ds^2 = - e^{2U} (dt + \omega_m dx^m)^2 + e^{-2 U} dx^m dx^m \ .
\ee
If in addition $\omega_m=0$ the metric becomes static. A natural guess for
the functions $e^{-2U}$ and $\omega_m$ follows from the requirement of
duality invariance. Note, there is no requirement for symplectic
invariance, only the duality subgroup represents an isometry.
In special geometry there are only two duality invariant expressions: 
the K\"ahler potential $K$ and the $U(1)$
K\"ahler connection $Q_{\mu}$. Below we show, that
\be170
\ba{rcccl}
e^{-2U}&=&e^{-K} & \equiv & i (\bar{X}^I F_I- X^I \bar{F}_I) \ , \\
{1 \over 2} e^{2U} \epsilon_{mnp} \partial_n \omega_p&=&Q_m & \equiv &
 {1 \over 2} e^{K}( \bar{F}_I \partial_m X^I - \bar{X}^I \partial_m F_I
 + c.c. ) \\
&&&=& { 1 \over 2} e^{2U} ( H_I \partial_m \tilde{H}^I - \tilde{H}^I
\partial_m H_I) 
\ea
\ee
gives a supersymmetric configuration. Similar expressions for $e^{-2U}$ and
$\omega_m$ have been suggested in \cite{be/ka}. Having in mind that the scalar 
fields are given by $z^A = X^A/X^0,$
we have expressed the solution completely in terms of the section
$(X^I , F_I)$, which is fixed by (\ref{010}). 
\subsection{Gravitino variation}
Before we start to show that (\ref{170}) provides a supersymmetric
configuration, we first collect some general formulae (our notations 
are given in the appendix). The Vielbeins for the IWP metric
are
\be175
 \ba{l}
 e_{\underline{0}}^{\ 0} = e^U \quad , \quad 
 e_{\underline{m}}^{\ 0} = e^U \, \omega_{\underline{m}} \quad , \quad 
 e_{\underline{m}}^{\ n} = e^{-U} \delta_{\underline{m}}^{\ n} \quad , \\
 e^{\underline{0}}_{\ 0} = e^{-U} \quad , \quad 
 e^{\underline{0}}_{\ m} = - e^U \, \omega_{\underline{m}} \quad , \quad 
 e^{\underline{m}}_{\ n} = e^{U} \delta^{\underline{m}}_{\ n} \ .
 \ea
\ee
For the anti-selfdual spin connections one obtains
\be190
\ba{l}
\omega_0^{-\,0i} = - {i \over 2} e^{2U} \bar{V}_i \quad , \quad
\omega_0^{-\, ij} = {1 \over 2} e^{2U} \epsilon^{ijk} \bar{V}_k, \\
\omega_m^{-\, 0i} = -{i \over 2} (e^{2U} \omega_m \bar{V}_i -
i \epsilon^{imk} \bar{V}_k), \\
\omega_m^{-\, ij} = {1 \over 2} (e^{2U} \omega_m  \epsilon^{ijk} 
\bar{V}_k -2i \delta_{m[i} \bar{V}_{j]}), 
\ea
\ee
with
\be192
\bar{V}_m = Q_m -i \, \partial_m U = Q_m - {i \over 2} \, \partial_m K.
\ee
Now we show that the configuration defined by (\ref{150}) --
(\ref{170}) with the holomorphic sections 
given by (\ref{010}) is supersymmetric.
We first transform the curved indices of the gauge fields as well as the
the $\gamma$-matrices into flat ones. The gauge fields are
defined by vanishing variations of (\ref{120}) and after taken into
account that $(F^I {^{\star}}G_I)_{curv.} =
\det|e_{\underline{\mu}}^{\ \mu}| (F^I {^{\star}}G_I)_{fl.}$ we get
\be230
\partial_{\mu} (\epsilon^{\mu\nu\rho\lambda} e^{-2U} 
G_{I\, \rho\lambda}) =0 \quad , \quad \partial_{\mu} 
(\epsilon^{\mu\nu\rho\lambda} e^{-2U} F^I_{\rho\lambda}) =0 \ .
\ee
As solution for the spatial components we obtain
\be240
F^I_{ij} = {1 \over 2} e^{2U} \epsilon_{ijp} \partial_p \tilde{H}^I
\quad , \quad G_{I\, ij} = {1 \over 2}e^{2U} \epsilon_{ijp} \partial_p
H_I \ .
\ee
Again the time components can be obtained by using the relation
(\ref{130}).
Inserting these fields into (\ref{044}) and using the anti-self-duality
condition (\ref{560}) we get for the graviphoton
field strength
\be250
\ba{l}
T^-_{ij} = {1 \over 2} e^{3U} \epsilon_{ijm} \left(F_I \partial_m
\tilde{H}^I - X^I \partial_m H_I\right) \ , \\
T^-_{0m} = {i \over 2} e^{3U} \left(F_I \partial_m \tilde{H}^I
 - X^I \partial_m H_I\right) \ .
\ea
\ee
Also, using the constraint (\ref{010}) and the relation (\ref{570}) one
finds
\be260
\partial_m e^{-2U} = (X^I +\bar{X}^I)\partial_m H_I -(F_I +\bar{F}_I)
\partial_m \tilde{H}^I
\ee
and in terms  of (\ref{170}) and (\ref{192}) one gets 
\be270
\bar{V}_m = i \, e^{2U} ( F_I \partial_m \tilde{H}^I - X^I \partial_m
H_I) = 2 \, e^{-U} T^-_{0m} \ .
\ee
Thus, we can write for the spin connections (\ref{190})
\be280
\ba{l}
\omega_0^{-\, 0i} = i e^U\, T^{-\, 0i} \quad , \quad
\omega_0^{-\, ij} = i e^U T^{-\, ij} \ ,\\ 
\omega_m^{-\, 0i} = i e^U \, \omega_m \, T^{- \, 0i} + e^{-U}
\epsilon^{imk} T_{k0}^- \ ,\\
\omega_m^{-\,ij} = i\, ( e^U \omega_m T^{- \, ij} + 2 \, e^{-U}
\delta_{m[i} T^-_{j]0} ) \ .
\ea
\ee
Transforming the curved index of $\gamma_{\mu}$ into a flat
one ($\gamma_{0} \rightarrow e_{\underline{0}}^{\ 0} \gamma_0
= e^U \gamma_0$ and $\gamma_m \rightarrow e_{\underline{m}}^{\ 0} \gamma_0 +
e_{\underline{m}}^{\ n} \gamma_n = e^U \omega_m \gamma_0 + e^{-U} \gamma_m$),
the time component of the gravitino supersymmetry variation 
(\ref{046}) gives
\be290
\delta \psi_{0 \alpha} = -{1 \over 4} e^U \, T^{-\, ab} 
\gamma_a \gamma_b \left(i \, \epsilon_{\alpha} + 
\gamma_0 \epsilon_{\alpha\beta} \epsilon^{\beta}\right) \ .
\ee
This variation vanishes if
\be300
\epsilon_{\alpha} = i\, \gamma_0 \epsilon_{\alpha \beta} \epsilon^{\beta}
\ .
\ee
As a consequence generically one half of supersymmetry is broken.
For static black holes with a non-singular horizon we know, that on
the horizon the complete $N=2$ supersymmetry is restored
\cite{G}. In order to investigate, whether also in the present case
supersymmetry restoration occurs at special points in space time
(note, there is no horizon in general), one has to look on the
variation of the gravitino field strength, i.e.\ at the  points of enhanced
supersymmetry $[\hat \nabla_{\mu} , \hat\nabla_{\nu}]\epsilon_{\alpha}$ 
should vanish
identically (if $\delta \psi_{\mu \, \alpha} = \hat \nabla_{\mu}
\epsilon_{\alpha}$). For a recent discussion see e.g.\ \cite{ka/ku}.

Next, using the relation (\ref{300}) and the chirality properties of the 
spinor (\ref{534}), we obtain from the spatial components of (\ref{046})
\be302
\delta \psi_{m \alpha} = \left(\partial_m +
 {1 \over 2}\, \partial_m U + i \, Q_m \right) \epsilon_{\alpha} \ .
\ee
Hence, in order to have non-trivial solutions the following
integrability constraint has to be fulfilled 
\be304
\partial_{[m} Q_{n]} =0
\ee
where $Q_m$ is the Kahler connection. 

\subsection{Gaugino variation}
This variation is given by
\be310
\delta \lambda^{\alpha A} = i \gamma^{\mu} \partial_{\mu} z^A 
\epsilon^{\alpha} + G^{-\, A}_{\rho \nu} \gamma^{\rho \nu} 
\epsilon^{\alpha\beta} \epsilon_{\beta}
\ee
with
$G^{-\, A}_{\rho \nu} = - g^{A\bar{B}} \,  
\bar f^I_{\bar B} \, \mbox{Im} {\cal N}_{IJ} F^{-\, J}_{\  \rho\nu}
$
where $g^{A \bar{B}}$ is the inverse K\"ahler metric and $\bar f^I_{\bar B}
= (\partial_{\bar{B}} + {1\over 2} \partial_{\bar{B}} K)
\bar{L}^I $ (see (\ref{020})).  

Multiplying (\ref{310}) with
$f^I_A$ and using the relation (\ref{038}) yields
\be330
f_A^I\delta \lambda^{\alpha A} = i  \gamma^{\mu} \partial_{\mu} z^A 
\left[(\partial_A + {1\over 2} \, \partial_A K) L^I \right] \epsilon^{\alpha} 
+ {1 \over 2} (F^{- \, I}_{\mu\nu} - i \bar{L}^I T^-_{\mu\nu}) 
\gamma^{\mu} \gamma^{\nu} \epsilon^{\alpha\beta} \epsilon_{\beta} \ .
\ee

Using the definition of the K\"ahler potential and $L^I = e^{K/2} X^I$ 
one obtains
\be340
\ba{l}
\partial_{\mu}z^A \partial_A e^{-K} = 
\bar X^J \partial_{\mu} H_J - \bar F_J \partial_{\mu} \tilde H^J \ , \\
\partial_{\mu}z^A \partial_A L^I = -{1 \over 2}\, e^{3/2 K} \, X^I\, 
\partial_{\mu}z^A \partial_A e^{-K} + \,e^{K/2} \,\partial_{\mu} X^I
\ .
\ea
\ee
Furthermore, in terms of (\ref{533}), (\ref{300}) and transforming 
the curved $\gamma$-indices into 
flat ones 
($\gamma^{\underline{m}} \partial_{\underline{m}} z^A 
\rightarrow e^U \gamma^m \partial_{\underline{m}} z^A$) one gets
\be350
f_A^I\delta \lambda^{\alpha A} =
\left( -2 e^U (\bar T_{0m}^- X^I - T_{0m}^- \bar X^I)  + i\, 
e^{2U} \partial_m X^I + 2i F^{- I}_{0m} \right) \gamma^m 
\epsilon^{\alpha}.
\ee
Using the relations (\ref{024}) and (\ref{020}) we find
\be360
\bar {\cal N}_{IJ} f_A^I\delta \lambda^{\alpha A} =
\left( -2 e^U (\bar T_{0m} F_J - T_{0m} \bar F_J)  + i \, 
e^{2U} \partial_m F_J + 2i G^-_{J\, 0m} \right) \gamma^m 
\epsilon^{\alpha} \ .
\ee
\mbox{From} (\ref{240}) it follows that
\be370
\ba{l}
F^{-I}_{0m} \equiv {1 \over 2}(F^{I}_{0m} - i\, ({^{\star}F^I})_{0m}) =
{1 \over 2}(F^I_{0m} + {i\over 2} \, e^{2 U} \partial_m \tilde H^I) \ , \\
G^{-}_{I\, 0m} \equiv {1 \over 2}(G_{I\, 0m} - i\, ({^{\star}G_I})_{0m}) =
{1 \over 2}(G_{I\, 0m} + {i \over 2} \, e^{2 U} \partial_m H_I)
\ea
\ee
and thus it can be easily seen that that the real parts of the expressions 
multiplying
the $\gamma^m\epsilon^{\alpha}$ term in (\ref{350}) and (\ref{360}) vanish. 
Consequently, $\delta
\lambda^{\alpha A}=0$ , because the matrix ${\cal N}_{IJ}$ has always
a non-trivial imaginary part (otherwise the gauge field part would be
pure topological) and the imaginary part is invertible ($\mbox{Im} {\cal N}
\cdot V =0 \leftrightarrow V=0$).

Thus, we have shown that the configuration (\ref{150}) -- (\ref{170})
with the holomorphic section constrained by (\ref{010})
defines a supersymmetric bosonic configuration, which breaks
generically one half of $N=2$ supersymmetry.

\section{Examples}
In this section we describe some explicit solutions. First we start with a
classification of harmonic functions, that yield different solutions,
like rotating black holes, Taub-NUT or Eguchi-Hanson instantons and
static solutions like extreme Reissner Nordstr{\o}m black holes.

As simplest example we start with pure supergravity.  Then, in terms
of the $STU$ model we will discuss various solutions for choices of the
harmonic functions. 
\subsection{Classification of solutions}
A general real harmonic function is given by the real or imaginary
part of a general complex harmonic function
\be440
H = \sum_n \left( C_n + {P_n \over \sqrt{(x - (x^0_n + i\, \gamma_n))^2 + 
(y- (y^0_n + i\, \beta_n))^2 + (z -(z^0_n  + i\, \alpha_n))^2}} \right) \ .
\ee
We have dropped here the symplectic index; $C_n$ and $P_n$ are complex
constants. Such a choice describes a multicenter solution, with the
constituents 
located at $(x^0_n , y^0_n , z^0_n)$ and $n$ counts all
centers. Note, that the integrability constraints (\ref{304}) will
give us restrictions for the parameters entering the harmonic
functions. In the following, we will ignore the multicenter case. The
singularity structure is determined by the parameters
($\gamma,\beta,\alpha$), e.g., if all are nontrivial, one obtains a
non-singular solution.  If, however, we set $\gamma=0$, we obtain a
solution with two singular points located at $y=z=0$ and $x = \pm
\sqrt{\beta^2 + \alpha^2}$.  More interesting for us are the cases of
the harmonic functions with $\gamma=\beta=0$ or $\alpha = \beta
=\gamma =0$. The first case has a ring singularity at $z=0$, $x^2 +
y^2 = \alpha^2$ and corresponds to rotating black holes. The second
case gives a spherically symmetric configuration and defines Taub-NUT
spaces, which are asymptotically not flat. In the multicenter case
when all the constants $C_n$ vanish one obtains a multi-instanton
solution, in the simplest case the Eguchi-Hanson instantons. Finally,
a vanishing Taub-NUT charge yields static solutions. Of course,
generically all the above mentioned solutions can appear in a
superposition.  For an analog classification of axion-dilaton black
holes see \cite{be/ka}.

\subsection{Pure supergravity}

The case of pure supergravity provides a toy model 
which illustrates our methods. 
The holomorphic 
prepotential of this theory is given by 
\be900
F=-{i\over4}(X^0)^2
\ee
where $X^0$ is the graviphoton complex scalar field. 
Eq (\ref{010}), gives in this case the following two equations
\be909
i(X^0-\bar X^0)=\tilde H^0,\qquad {1\over2}(X^0+\bar X^0)=H_0
\ee
which implies that $X^0$ is set by a complex harmonic function
\be919
X^0=H_0-{i\over2}{\tilde H}^0
\ee
and the solution is then given by
\be929
ds^2=- {1\over X^0\bar X^0}(dt+\omega_m dx^m)^2+{X^0\bar X^0}dx^2 \ .
\ee
If we choose for $X^0$ the complex harmonic function
\be453
X^0=1+{{M+iN}\over r}
\ee
where $M$ and $N$ are real constants and $r$ is the radial 
distance in spherical 
coordinates. Here, $\omega$ has only one component 
$\omega_\phi$ which is given by the following equation
\be979
{1\over r^2\sin\theta}\partial_\theta\omega_\phi=
H_0\partial_r\tilde H^0-\tilde H^0\partial_r H_0
\ee
where
\be956
H_0=1+{M\over r}; \qquad {\tilde H}^0=-{2N\over r} \ .
\ee
Therefore one obtains the following metric
\be934
ds^2=-{1 \over \left[(1+{M\over r})^2+{N^2\over r^2}\right]}
\left(dt+2N\cos\theta d\phi\right)^2 +
\left[(1+{M\over r})^2+{N^2\over r^2})\right]d\Omega^2
\ee
where $d\Omega^2=(dr^2+r^2d\theta^2+r^2\sin^2\theta d\phi^2)$.
This is the extreme electromagnetic generalization of the Taub-Nut metric
considered by Hawking and Hartle \cite{hh}.

Next, we choose a different harmonic function for $X^0$, 
\be999
X^0=1+{m\over\sqrt{x^2+y^2+(z-i\alpha)^2}}
\ee
In this case, it is convenient to use the so-called oblate
spheroidal coordinates, which are defined by
\be450
x+ i y = \sqrt{ r^2 + \alpha^2}\sin\theta e^{ i \phi};\qquad 
z = r \cos \theta,
\ee
where the flat spatial
metric becomes
\be451
dx^m dx^m = {r^2 + \alpha^2 \cos^2 \theta \over r^2 + \alpha^2}
\, dr^2 + (r^2 + \alpha^2 \cos^2 \theta) \, d\theta^2 + (r^2 + \alpha^2)
\sin^2\theta \, d\phi^2
\ee
and
\be988
X^0= 1+{m(r+i\alpha\cos\theta)\over r^2+\alpha^2\cos^2\theta}
\ee 
In this case we obtain for $\omega_\phi$ and $e^{-K}$ the following
\be975
e^{-K}={(r+m)^2+\alpha^2\cos^2\theta\over r^2+\alpha^2\cos^2\theta}, 
\qquad
\omega_\phi={(2mr+m^2)\alpha\sin^2\theta\over r^2+\alpha^2\cos^2\theta}.
\ee
Thus one obtains the BPS-saturated 
Kerr-Newman metric \cite{is/wi}.

\subsection{Solution for the symplectic section of the $STU$ model}
The classical $STU$ model is given by 
the prepotential
\be400
F_{cl.}(X) = - {X^1 X^2 X^3  \over X^0}.
\ee
For a supersymmetric configuration, $X^I$ are given as solutions of
({\ref{010}).  Since these equations are similar to the stabilisation
equations after replacing $(\bar ZL^I, \bar Z M_I)$ with $(X^I, F_I)$ and
  the charges by harmonic functions, 
we can take known results from double extreme black hole solutions and
replace the charges by harmonic functions. For the model at
hand the double extreme solution has been discussed in \cite{be/ka1}
\footnote{Note, that in these references the solution
is given in a different symplectic basis, where $X^0=1$.}.
If one first defines the ``symmetrized $\epsilon$-tensor'' 
with non-vanishing components given by
\be410
1= d_{123} = d^{123} = d_{213} = (\mbox{all permutations})
\ee
the solution for the $STU$ model can be written as 
\be420
e^{-4U} = e^{-2K} = - (\tilde H^I H_I )^2 
+ ( d_{ABC}\tilde H^B \tilde H^C d^{ADE} H_D H_E )
 + 4 \tilde H^0 H_1 H_2 H_3 - 4 H_0 \tilde H^1 \tilde H^2 \tilde H^3
\ee
and the scalar fields are 
\be430
z^A = {X^A \over X^0} = { \left(2 \tilde H^A H_A  - (\tilde H^I  H_I)  
\right) - i \,
e^{-2U} \over (d_{ABC} \tilde H^B \tilde H^C + 2 \tilde H^0 H_A)} 
\ee
with no summation over the index $A=1,2,3$ but over $I=0,1, 2,3$. 
These equations define the complete symplectic coordinates $X^I$; 
$X^0$ can be obtained by inserting $z^A$ into the 
K\"ahler potential (\ref{170}).

\subsection{Rotating black holes}

By choosing different harmonic functions $(\tilde H^I, H_I)$ in
(\ref{440}) one can obtain different types of solutions. For exmple,
as mentioned earlier, rotating black holes could be obtained if we
choose for our solution harmonic functions with $\gamma=\beta=0.$ This
provides an axial-symmetric configuration with the $z$-direction as
rotational axis.  Here it is more convenient to go to the appropriate
coordinate system. As for the pure gravity case we use the spheroidal
coordinates, in terms of which, the general (one-center)
complex harmonic function is given by
\be452
H = C + {Q\over \sqrt{x^2 + y^2 + (z - i\alpha)^2}}=  
C + {Q( r + i\alpha \cos \theta)\over r^2 + \alpha^2 \cos^2\theta}
\ee
and therefore the ring singularity is at $r = \cos\theta =0$.

Next, rotating black holes are asymptotically flat, which implies
\be460
e^{2U} \rightarrow 1 \qquad \mbox{and} \qquad \omega_m \rightarrow 0 
\quad, \quad \mbox{for} \quad r \rightarrow \infty \ .
\ee
The first condition in the above equation puts constraints on the
additive constants in the harmonic functions. More interesting is the
second part. The axial symmetry requires that nothing depends on the
coordinate $\phi.$ Transforming (\ref{170}), which defines $\omega_m,$
into spheroidal coordinates, we obtain \footnote{ As consequence of
the axial symmetry $\omega_m$ has only a $\phi$ component and the
determinant of the 3-d metric is $\sqrt{g} =(r^2 + \alpha^2
\cos^2\theta)\, \sin\theta$.}
\be470
\ba{rcl}
{1 \over (r^2 + \alpha^2) \sin\theta} 
\; \partial_{\theta} \, \omega_{\phi}&=&H_I \partial_r \tilde H^I - 
\tilde H^I \partial_r H_I \ ,\\
- {1 \over \sin\theta}  \; \partial_{r} \, \omega_{\phi}&=& 
H_I \partial_{\theta} \tilde H^I - \tilde H^I \partial_{\theta} H_I \ . 
\ea
\ee
Therefore, in order to have asymptotic flat solutions we have to
choose the harmonic functions in such a way that the rhs for 
the first equation 
in (\ref{470}) vanishes
faster than $1/r^2$ and at least as $1/r^2$ for
the second equation. This is possible if one takes the following Ansatz for
$(\tilde H^I , H_I)$
\be480
\ba{l}
\tilde H^0 = \mbox{Im} \left( {m^0 \over \sqrt{x^2 + y^2 + 
(z - i \, \alpha)^2}} \right) = {m^0 \alpha \cos \theta \over r^2 +
\alpha^2 \cos^2 \theta} \ , \\
\tilde H^A = \mbox{Re} \left( h^A  + {p^A \over \sqrt{x^2 + y^2 + 
(z - i\, \alpha)^2}} \right) =  h^A + {p^A \over R} \ , \\
H_0 =\mbox{Re} \left( h_0 + { q_0 \over \sqrt{x^2 + y^2 + 
(z - i\, \alpha)^2}} \right) =  h_0 + {q_0 \over R} \ , \\
H_A = \mbox{Im} \left( {m_A \over \sqrt{x^2 + y^2 + 
(z - i\, \alpha)^2}} \right) = {m_A \alpha \cos \theta \over r^2 +
\alpha^2 \cos^2 \theta} 
\ea
\ee
where we introduced $R= (r^2 + \alpha^2 \cos^2 \theta)/r$.
Furthermore, in order to get an asymptotic Minkowski space
($U \rightarrow 0$) we have the constraint
\be482
- 4 h_0 h^1 h^2 h^3 = 1 \ ,
\ee
i.e.\ we have to choose one of the $h's$ to be negative, e.g.\ $h_0 < 0$.
It can also be seen as a fixing of $h_0$ and the additional three parameter
$h^A$ parameterize the scalar fields at infinity
\be490
z^A_{\infty} = - i \, { 2 h_0 h^A \over \sqrt{ - 4 h_0 h^1 h^2 h^3}} \ .
\ee
Thus asymptotically, all axions vanish.  It is also useful to
determine our section at infinity. From (\ref{010}) one sees that
$X^0_{\infty}$ has to be real and thus $X^A_{\infty}$ are pure
imaginary ($z^A_{\infty}$ is pure imaginary) and one obtains
\be600
X^0_{\infty} = \sqrt{-  h^1 h^2 h^3 \over 4 h_0} \qquad , \qquad 
X^A_{\infty} = - i\, {h^A \over 2} \ .
\ee
Inserting this expression into the prepotential we find
\be610
F_0^{\infty} = - i \, {h_0 \over 2}  \qquad , \qquad 
F_A^{\infty} = { {1 \over 2} h_0\, d_{ABC} h^B h^C \over 
\sqrt{- 4 h_0 h^1 h^2 h^3}} \ .
\ee
Next, we need to ensure that  the integrability constraint (\ref{304}) is 
satisfied. This constraint in terms of the harmonic functions takes the 
form 
\be612 
2 \partial_{[n} Q_{m]} = \partial_{[n} \left( e^{2U} (H_I \partial_{m]} 
\tilde{H}^I - \tilde{H}^I \partial_{m]} H_I) \right)=0 
\ee
which is solved if
\be614
\partial_{[n} \tilde H^I \partial_{m]} H_J = 0 \ .
\ee
For our harmonic functions we have
\be616
\ba{rcl}
\partial_n \tilde H^0 \partial_m H_0 & \sim & m^0 q_0 \; 
\mbox{Im}( \partial_n {1 \over r_1}\, \partial_m {1 \over r_1}) \ ,\\
\partial_n \tilde H^A \partial_m H_B & \sim & p^A m_B \;
\mbox{Im}( \partial_n {1 \over r_1} \, \partial_m {1 \over r_1}) \ , \\
\partial_n \tilde H^0 \partial_m H_A & \sim & m^0 m_B \;
\partial_n \mbox{Im}({1 \over r_1})\,
\partial_m \mbox{Im}({1 \over r_1}) \ ,\\
\partial_n \tilde H^A \partial_m H_0 & \sim & p^A q_0\;
\partial_n \mbox{Re}({1 \over r_1}) \, \partial_m \mbox{Re}({1 \over r_1}) 
\ea
\ee
where $r_1 = \sqrt{x^2 + y^2 + (z - i \, \alpha)^2}$.
So, after antisymmetrizing  the derivatives, the condition (\ref{614})
is identically fulfilled.

In order to give the parameters that enter the harmonic functions
a physical meaning we go to the asympotic flat region.

First from the asymptotic behavior of the metric we get the mass and the
angular momentum
\be620
\ba{l}
- g_{00} = e^{2U} = 1 - {2M \over r} \, \pm .. \ = 1 +  
{2 \, (q_0 h^1 h^2 h^3 + 1/2 \, h_0 p^A \, 
d_{ABC}h^B h^C) \over r} \, \pm .. \ ,\\
- g_{0\phi} = e^{2U} \omega_{\phi} = {2 J \, \sin^2\theta \over r} \, \pm .. \ 
= {2 (h^A m_A - h_0 m^0) \, \alpha \, \sin^2\theta \over r} \, \pm .. \ .
\ea
\ee
Thus, the mass is positve only, if some of the parameters are negative
and we will assume that $H_0< 0 \ \forall r$, i.e.\ $h_0, q_0 < 0$.
By using of (\ref{482}) the mass and angular momentum of our 
solution is given by 
\be630
M =  |Z| = X^0_{\infty} q_0 - F_A^{\infty} p^A \quad , \quad 
J = (h_0 m^0 - h^A m_A ) \, \alpha \ 
\ee
where $Z$ is the central charge and $q_0, h_0 <0$.
Finally, for the gauge fields we get asymptotically 
\be640
\ba{lcl}
(^{\star}F^0)_{0r} = 2\, {m^0 \alpha \cos\theta \over r^3}  
 \pm .. & \quad , \quad  & 
(^{\star}F^A)_{0r} = {p^A \over r^2}  \pm ..  \quad , \\
(^{\star}G_{0})_{0r} = {q_0 \over r^2}  \pm .. & \quad ,  \quad &
(^{\star}G_A)_{0r} = 2\, {m_A \alpha \cos\theta \over r^3} \pm .. \quad . 
\ea
\ee
Therefore, the black hole carries 3 magnetic ($p^A$),
one electric ($q_0$) charges;  as well as 3 electric  
(${1 \over 2} m_A \, \alpha$)
and one magnetic (${1 \over 2} m^0\, \alpha$) dipole momenta.

These expressions suggest now the following interpretation.  The
rotating black hole is a bound state that consist of 4 constituents
each determined by two harmonic functions ($\tilde H^I , H_I$), i.e.\
3 magnetically charged and one electrically charged.  For every
consituent the gyromagnetic ratio is 1. E.g.\ extracting the $(\tilde
H^1 , H_1)$ part and making all others trivial, 
one finds for the dipol moment
\be642
\mu_1 = {1 \over 2} \, g \, J \, {p^1 \over M}
\ee
with $g=1$. This is an expected result, since every constituent
is a so-called $a=\sqrt{3}$ black hole, which should be equivalent to
Kaluza-Klein states; for a recent review see \cite{du/ra}.

Already the BPS bound for the mass (\ref{630}) suggest this bound
state interpretation. For our solution here, the central charge on the
rhs.\ is real, i.e.\ $M=|Z|$ and the total mass is a direct sum of the
masses of their consituents. Also we note that the angular momentum does not
enter into the BPS bound. This has a serious consequence: the black hole
exhibits a naked singularity; there is no horizon hiding the ring
singularity\footnote{So, strongly speeking these objects are not really
black, but for convenience we will use the name black hole
for any asymptotically flat 0-brane (no spatial translation isometries).}. 
Different speculations have been made in the literature 
to overcome this
shortcoming. First, in forming macroscopic black holes one could
argue that the starting point is a non-extreme black hole, which will then 
evaporate until it becomes extremal. In this proccess it can
also lose angular momentum and reaches the minimal mass as a static
supersymmetric black hole, see e.g.\ \cite{cv}. Other arguments suggests that
it could be the wrong way in giving a black hole (or general any
kind of 0-branes) angular momentum. Alternatively, the angular
momentum could come from fermionic hairs \cite{du}. Or, one could
construct a rotating black hole solution, which asymptotically does not
differ, but has at the core a much milder singularity and in addition
has a Regge-bounded angular momentum \cite{da/ga}.

One can also speculate about a different point of view. As we will
discuss later there is a limit in which the ring singularity could be 
hidden, namely in the massless case. In this limit an additional
singularity appears at a finite radius - where in the internal space
an expected topology change takes place. By doing this one avoids the
naked singularity, but obtains a massless black hole singularity
which is not well understood either. On the other hand this opens the
interpretation that rotating black holes can consistently exist only
as massless gauge bosons. So far this point of view is rather
speculative and we will leave it for further investigations.


\subsection{Taub-NUT  and Eguchi-Hanson instantons}
As a next example we will 
discuss the case $\alpha=\beta=\gamma=0$, 
but $Q_m \neq 0$, which includes Taub-NUT spaces and Eguchi-Hanson instantons.
We take for the harmonic functions
\be680
\ba{l}
\tilde H^0 = \tilde h^0 + {m^0 \over r }  \quad , \quad
\tilde H^A = \tilde h^A  + {m^A \over r} \ , \\
H_0 = h_0 + {n_0 \over r} \quad , \quad
H_A = h_A + {n_A \over r} 
\ea
\ee
where $r$ is now the standard 3-d radius. In this case $\omega_m$
is defined by
\be690
{1 \over r^2\sin\theta}\partial_{\theta} 
\omega_{\phi}= -{(  h_I m^I -  \tilde h^I n_I ) \over r^2} \ ,
\ee
As solution one finds 
\be700
\omega_{\phi} = 2 l \cos\theta \qquad , \qquad l = 
{1 \over 2} (h_I m^I - \tilde h^I n_I)  
\ee
where $l$ as Taub-NUT charge. Also, in this case the integrability
constraint (\ref{304}) is identically fulfilled.

This solution is consistently defined only after removing the
Dirac-singularity, which appears at $\theta = \pi$ or 0.  As usual,
one has to introduce different coordinate patches for the north and
south hemisphere without a singularity.  To do this consistently one
has to assume that the time is periodic ($t \simeq t + 8\pi n$). Also,
since $\omega_m$ does not vanish at infinity, this solution is not
asymptotically flat.  As Euclidean solutions the Taub-NUT space is one
example for a gravitational instanton.

To find further solutions, we generalize the harmonic
functions to the multi-center case
\be710
\ba{l}
\tilde H^0 = \tilde h^0 + m^0 \sum_i^k {1  \over r_i }\quad , \quad
\tilde H^A = \tilde h^A  + m^A \sum_i^k {1 \over r_i} \ , \\
H_0 = h_0 + n_0 \sum_i^k {1 \over r_i} \quad , \quad
H_A = h_A + n_A \sum_i^k {1 \over r_i} 
\ea
\ee
with $r_i^2 = |\vec{x} - \vec{x_i}|^2$.  We have taken the same charge
for every center in order to satisfy the integrability constraint
(\ref{304}).  This is a generalization of the $(k-1)$-multi-instanton
configuration; for non-vanishing $h's$ related to multicenter Taub-NUT
and for vanishing $h's$ to Eguchi-Hanson instantons.  As for the
rotating case, this solution can be seen as a bound state where every
constituent is represented by one harmonic function.  One has however
to keep in mind that these solutions here are generalisations in the
sense, that they couple to further scalars as well as gauge fields.  In
the standard form, these instantons have a selfdual curvature tensor
\cite{gi/ha} and for vanishing constant part ($h's$) it is flat space
time for $k=1$ , for $k=2$ it is an Eguchi-Hanson instanton and in
general it is a $(k-1)$-multi instanton configuration (see also
\cite{eg/ha}).

\subsection{Static limit and quantum corrections}

Finally we want to discuss the static limit, which 
is defined by
$Q_m=\omega_m=0$. Going back to the rotating black hole solution
this limit can be obtained by setting $\alpha=0$, i.e.\ vanishing
angular momentum. The four harmonic
functions are
\be730
H_0 = h_0 + {q_0 \over r} \qquad , \qquad 
\tilde H^A = h^A + { p^A  \over r}
\ee
with $A=1, 2, 3$ and the metric is given by
\be740
ds^2 = -e^{2U} dt^2 + e^{-2U} dx^m dx^m \quad  , \quad 
e^{-2U} = \sqrt{- 4 H_0 \tilde H^1 \tilde H^2 \tilde H^3}
\ee
the scalar fields are
\be750
z^A = - i\, {2 H_0 \tilde H^A \over 
\sqrt{- 4 H_0 \tilde H^1 \tilde H^2 \tilde H^3}} \quad , \quad S = -i\, z^1 
\quad , \quad T = -i \, z^2 \quad , \quad  U = -i \,z^3
\ee
and the gauge fields are defined in (\ref{150}).  Again one harmonic
function has to be negative definite, e.g.\ $H_0<0 \ \forall r$.  We
see, that all axionic parts in the scalar fields vanished in the
static limit, which is a consequence of the simple choice for the
harmonic functions and has nothing to do with static limit. The
axion-free solutions are especially interesting since they allow a
clear brane interpretation. The role of the axions in the brane
picture is not yet completely understood.  Perhaps they are important
for an understanding of tilted branes \cite{be/cv}.  On the type IIA
side the model at hand has the interpretation of an intersection of
three 4-branes and a 0-brane, or in the $M$-brane picture, it is an
intersection of three 5-branes with a boost along the common string.
The 11-d solution is given by \cite{pa/to}
\be760
\ba{l}
ds^2_{11d} = {1 \over (H^1 H^2 H^3)^{1/3} }
\left[ dudv + H_0 du^2 + H^A d\chi_A + H^1 H^2 H^3 d\vec{x} d\vec{x} 
\right] \ ,\\
F = {^{\star}}\left( \epsilon^{ABC} d(1/H^A) \wedge d\chi_B \wedge d\chi_C 
\wedge du \wedge dv \right) 
\ea
\ee
where $\chi_A$ are three 2-d line elements and $u,v = z \pm
t$ where $z$ is the common string direction.  This configuration is
compactified first to $D=5$ by wrapping the 5-branes around 4-cycles
of the torus. As consequence we get the magnetic string, which,
when wrapped around the $5th$ direction, yields the black hole solution
(\ref{740}). Equivalently, after compactification one can see that the 4-d
black hole as consisting of 4 ``elementary black holes'', the
so-called $a=\sqrt{3}$ black holes \cite{ra}.

The mass of this solution is (using the relation
(\ref{482}))
\be820
M= - q_0 h^1 h^2 h^3  + {p^A  d_{ABC} h^B h^C \over 8 h^1 h^2 h^3}
\ee
which coincides with the mass of the rotating black hole (\ref{620}),
note $q_0<0$.
The $h's$ define the vev's of the scalar fields and for certain
values the black hole mass becomes extremal (minimal). This
defines the double extreme limit \cite{ka/sh}. Calculating 
$\partial/\partial h^A
M =0$ and using the ansatz
\be830
h_0 = {q_0 \over c} \qquad , \qquad h^A = {p^A \over c}
\ee
one finds
\be832
c^4 = - 4 q_0 p^1 p^2 p^3 \ .
\ee
This gives the minimal mass as long as $q_0<0, p^A>0$.
Plugging these values into the scalar fields (\ref{750}) 
the space-time dependence drops out completely, they are constant
and given only by the conserved charges. In this case the mass 
coincides with the Bekenstein-Hawking entropy \cite{bh}
that measures the area of the horizon at $r=0$ 
\be810
{\cal S} = {A \over 4} = \pi \left( r^2 e^{-2U} \right)_{r=0} \ .
\ee

On the other hand, allowing different signs between the $h's$ and the
charges one finds massless black holes \cite{be1}. A simple example is
given by \cite{cv2}
\be812
e^{-2U} = \sqrt{(1+ {q \over r})(1 - {q \over r})(1 + {p\over r})
(1 - {p \over r})}
\ee
i.e.\ we took $p^1 = - q_0 = q$ , $p^3 = - p^2 =p$ and all $h's$ are
trivial. Since there is no $1/r$ term, the mass vanish
identical. However, as consequence, additional singularities appear
at $r= p,q$. From the brane picture these negative charges are defining
anti-branes, or after compactification, the bound states
contains constituents with negative mass. As, e.g., recently discussed
in \cite{wa}, we have to expect that these singular points signal a
topology change (see also below). 

These singularities create a gravitational potential that is repulsive
to all matter (``anti-gravity'') \cite{li/ka}. This is a welcomed
feature because, as consequence, this massless black hole  at rest
is not stable. Instead any further matter will accelerate it untill
the speed of light is reached \cite{or}.  Also, as speculated earlier,
the potential barrier could hide the naked singularity of rotating
black holes, with the consequence that rotating supersymmetric black
holes (or 0-branes) have to be massless and cannot be at
rest. We have, however, to admit that these singular points are not
yet well understood in the black hole picture. 

At these points a K\"ahler class modulus (\ref{750}) vanishes
indicating a vanishing 4-cycle. In our example $z^1$ and $z^3$ vanish
at $r=q$ or $r=p$ resp.  and near these points our classical solution
breaks down, it cannot be trusted anymore.  It is well-defined as long
as all scalar fields $z^A$ are large enough, which is equivalent to
the requirement $|H_0| \gg H^A \gg 1$, $\forall r$. The first
inequality ensures that
all 4-cycles have everywhere large radii whereas the last one ensures,
that $\alpha'$ corrections (e.g.\ $R^2$ terms) can be neglected.
Since near these singular points the solution is describable by a
conformal exact model \cite{ho/ts}, \cite{be1} it is unlikely that
these singularities can be cured by $\alpha'$ corrections (higher
derivative terms).  We have however to expect that near a vanishing
4-cycle quantum corrections on the heterotic side or worldsheet
instanton corrections on the type II side becomes important.  Let us
discuss the structure of these corrections on the type II side.

On the type II side (heterotic) quantum corrections are encoded in the
topological structure of the Calabi-Yau (CY) threefold upon which we have
to compacitfy.  The complete prepotential is given by \cite{c}
\be770
F(X^A) = (X^0)^2 \left[ -{1\over 6} C_{ABC} z^A z^B z^C - 
{i \chi \zeta (3) \over
2 (2 \pi)^3} + {i \over (2 \pi)^3} \sum_{d_1,..d_n} n^r_{d_1, .. , d_n}
Li_3(e^{2 \pi i \, d_A z^A} ) \right] \ . 
\ee
Here $C_{ABC}$ are the classical intersection numbers, $\chi$ is the
Euler number of the CY manifold, $n^r_{d_1, .. , d_n}$ are the numbers
of genus zero rational curves (instanton numbers) and $d_A$ are the
degrees of these curves. These numbers are determined by the
CY-threefold of the compactified effective Type II string
theory. For this prepotential we have now to solve the equations
(\ref{010}).  In general this is hopeless. But we are mainly
interested in corrections around a vanishing 4-cycle, say $z^3$. 

Again, in solving the algebraic constraint (\ref{010}) we can adopt
results for the double extreme case and find as solution for
the symplectic section \cite{be/ga}
\be780
X^0 = {\lambda \over 2} \quad , \quad  X^A = - i\, { H^A \over 2} 
\qquad \mbox{and} \qquad z^A = - i \, {H^A \over \lambda}  \ .
\ee
The parameter $\lambda$ is fixed by the equation $F_0 - \bar F_0 = -i H_0$ 
and expanding the solution around $z^3 \simeq H^3 \simeq 0$ one finds 
\be790
\lambda =  \pm \sqrt{- {1 \over 6} C_{ABC} H^A H^B H^C \over H_0} +
\sum_{d_3} n^r_{d_3} {(d_3 H^3)^{2} \over 8 \pi H_0} \pm .. \ .
\ee
Inserting this into $e^{-2U}$ yields
\be800
e^{-2U} =  \sqrt{- H_0 {1 \over 6} C_{ABC} H^A H^B H^C} -
\frac{1}{8} \sum_{d_3} n^r_{d_3}  (d_3 H^3)^2 \log 
\left(    \frac{ (d_3H^3)^{2} H_0}{ -\frac{1}{6}C_{ABC}H^A H^B H^C}
\right) \pm .. \ \ .
\ee
The black hole (\ref{740}) includes now all perturbative and
non-perturbative corrections in the neighbourhood of a vanishing
4-cycle. Following the procedure described in \cite{be/ga} one could
easely calculate further corrections.  But the general structure is
the same, the black hole receives polynomial and logarithmic corrections.
Also in the entropy, the instanton corrections yield additional
logarithmic corrections. For a microscopic discsussion 
see also the second ref.\ of \cite{be/ga}.

If the K\"ahler potential behaves smooth at this point ($z^3 \simeq
0$) , the corrections regularize the metric. But the system can
undergo a phase transition, if derivatives of the K\"ahler potential,
like the K\"ahler metric or gauge couplings, behave non-smoothly.  We
have to expect different types of phase transitions, mild forms like
the flop transisition as recently discussed in \cite{ch/ka}, but also
drastic ones like conifold transitions. Physically, at these points an
internal cycle of the CY vanishes and ``beyond'' this point a 
new,
topologically different cycle emerges.

We will leave further discussions of these interesting phenomena for
future investigations.  Especially, it would be very interesting to
understand in our framework the connection to the massless black holes 
described in the
context of the resolution of conifold singularities \cite{st}.

\section{Conclusions}
The stabilization equations for the double extreme black holes has
been a very fruitful framework in addressing quantum corrections
for supersymmetric black holes. They allow to calculate the entropy for
all black holes with non-singular horizons and describe black
hole solutions that have been stabilized at their minimal mass.

In this paper we generalized this framework to incorporate 
general solutions of
$N=2$ supergravity, which include not only black holes (non-singular
and singular ones) but also Taub-NUT and Eguchi-Hanson instantons.

Our starting point was a general Ansatz for a stationary metric, which
should be completely determined by duality invariant quantities,
namely by the K\"ahler potential and the $U(1)$ K\"ahler connection (see
section 3.1).  Next, we showed that solutions of our constraint
equations (\ref{010}) define supersymmetric configurations (section
3.2 and 3.3). Note, although these configurations are supersymmetric
(vanishing gravitino and gaugino variation), they do {\bf not} solve
the equations of motion.  They are off-shell supersymmetric
configurations defined by arbitrary functions $(\tilde H^I ,
H_I)$. However, the gauge field equations and Bianchi identities are
fulfilled only, if the section $(\tilde H^I , H_I)$ consists of
harmonic functions.  As result, we get an on-shell supersymmetric
solution. Since the supersymmetry variations are only first order
differential equations, the harmonic property of the functions can
never be important. Therefore, equations (\ref{010}) can be seen as
defining a consistent supersymmetric truncation to a bosonic theory,
i.e.\ vanishing gravitino and gaugino variations. As usual, this
truncation breaks one half of supersymmetry.

These equations have also a nice physical meaning, they make the
interplay between the internal and external space obvious. Whenever, one
approaches a singular/special point in space time one moves in the internal
space to singular/special points too and vice versa. One cannot separate
the internal and external space. It is the gravity, that unifies both parts.

In a second part, we discussed explicit solutions, starting with the pure
supergravity case. Different types of solutions are related to different
choices of harmonic functions. We discussed some examples for the $STU$
model, like rotating black holes, Taub-NUT and Eguchi-Hanson instantons
and generalization for the Reissner-Nordstr{\o}m black hole.

Since our approach is not restricted to special types of solutions, it
can be used to investigate singularities. For a static black hole we have
discussed an example. The classical solution becomes singular at points
where a 4-cycle vanishes, around which a 5-branes is wrapped. However, if
one takes into account all perturbative and non-perturbative corrections
and expand the solution around this point, one finds that the black hole
solution behaves non-singular as long as the K\"ahler potential remains
non-singular. In addition to polynomial corrections, the black hole
receives logarithmic corrections. Although, for a wide range of models the
K\"ahler potential stays non-singular, their derivatives, like K\"ahler
metric or gauge couplings, will not be smooth indicating a phase
transition. 

Of course, all examples can only be seen as a starting point for further
investigations. The framework described in the paper could provide a
solid basis for addressing 
questions related to singular points of the
Calabi-Yau, e.g.\ different types of phase transitions. For the rotating
black holes one could ask, whether quantum corrections could remove/hide
the naked singularity in the supersymmetric limit.

\newpage

\noindent
{\Large \bf Appendix: conventions and notation}

For the metric we are using the signature $(-+++)$ and for the indices
we take: $\mu, \nu .. = 0,1,2,3$, whereas $m,n,\cdots = 1,2,3$.  Where
confusions can arise, we will denote the underlined indices
$\underline{\mu} , \underline{\nu}$ as curved ones and $a,b,\cdots =
0,1,2,3$ as flat ones.  Antisymmetrized indices are defined by: $[ab]
= {1 \over2} (ab -ba)$.

The spin connections is
\be510
\omega_{\mu}^{ab} = 2 e^{\nu[a} \partial_{[\mu} e_{\nu]}^{\ b]}
 - e^{\rho a} e^{\sigma b} e_{\mu p } \partial_{[\rho} e_{\sigma]}^{\ p} \ .
\ee
We define (anti) selfdual components like
\be520
F^{\pm}_{ab} = {1 \over 2}(F_{ab}  \pm i\, {^{\star}F_{ab}})
\ee
with
\be530
{^{\star}F^{ab}} = {1 \over 2} \epsilon^{abcd}F_{cd}
\ee
and $\epsilon^{0123} = 1 = - \epsilon_{0123}$ ($ ^{\star\star}F= -F$).

For $\gamma$ matrices we are using the relation
\be532
\gamma^a \gamma^b = - \eta^{ab} + {i \over 2} \gamma_5 \epsilon^{abcd}
\gamma_c \gamma_d
\ee
with $\gamma_5 = -i \gamma_0 \gamma_1 \gamma_2 \gamma_3$ ($\gamma_5^2 =1$).
Using these definitions we find for any antiselfdual tensor
the identity
\be533
T^{-\, ab} \gamma_a \gamma_b = 2(1 - \gamma_5) T^{-\, 0m} \gamma_0 \gamma_m
\ee
and
\be560
T^-_{mn} = - i\, \epsilon_{mnp} T^-_{0p} \ .
\ee
For the chiral spinors (gaugino and gravitino) we have
\be534
\gamma_5 \pmatrix{\lambda^{\alpha} \cr \psi_{\alpha}} = -
 \pmatrix{\lambda^{\alpha} \cr \psi_{\alpha}} \quad , \quad
\gamma_5 \pmatrix{\lambda_{\alpha} \cr \psi^{\alpha}} = 
\pmatrix{\lambda_{\alpha} \cr \psi^{\alpha}} \quad . 
\ee
Note, that our conventions slightly differ from the ones 
used
in \cite{an/be}. Our signature and $\epsilon$-tensor definition
has the consequence that also the spinor chiralities are opposite.

\bigskip
\bigskip

{\large \bf Acknowledgements}

\medskip

\noindent
We would like to thank I.\ Gaida, S.\ Mahapatra,  A. Van Proeyen,
 and especially 
T.\ Mohaupt for interesting discussions.  The work is supported by 
the Deutsche Forschungsgemeinschaft (DFG) and by the European Commision TMR
programme ERBFMRX-CT96-0045.  W.A.S is partially supported by
DESY-Zeuthen.

\end{document}